\documentclass[sn-basic]{sn-jnl}

\usepackage{graphicx}%
\usepackage{multirow}%
\usepackage{amsmath,amssymb,amsfonts}%
\usepackage{amsthm}%
\usepackage{mathrsfs}%
\usepackage[title]{appendix}%
\usepackage{xcolor}%
\usepackage{textcomp}%
\usepackage{manyfoot}%
\usepackage{booktabs}%
\usepackage{algorithm}%
\usepackage{algorithmicx}%
\usepackage{algpseudocode}%
\usepackage{listings}%
\usepackage{adjustbox}
\graphicspath{{fig/}}
\newcommand{\rev}[1]{\textcolor{black}{#1}}

\usepackage{CJKutf8}

\theoremstyle{thmstyleone}%
\theoremstyle{thmstyletwo}%

\theoremstyle{thmstylethree}%

\begin{document}

\title[Article Title]{Mapping Urban Villages in China: Progress and Challenges}


\author*[1]{\fnm{Rui} \sur{Cao}} \email{ruicao@hkust-gz.edu.cn}
\author[2]{\fnm{Wei} \sur{Tu}}
\author[3]{\fnm{Dongsheng} \sur{Chen}}
\author[1,4]{\fnm{Wenyu} \sur{Zhang}}

\affil[1]{\orgdiv{Thrust of Urban Governance and Design, Society Hub}, \orgname{The Hong Kong University of Science and Technology (Guangzhou)}, \orgaddress{\city{Guangzhou}, \country{China}}}
\affil[2]{\orgdiv{MNR Key Laboratory for Geo-Environmental Monitoring of Great Bay Area, Guangdong Key Laboratory of Urban Informatics, Shenzhen Key Laboratory of Spatial Smart Sensing and Services, Department of Urban Informatics}, \orgname{Shenzhen University}, \orgaddress{\city{Shenzhen}, \country{China}}}
\affil[3]{\orgdiv{Chair of Cartography and Visual Analytics}, \orgname{Technical University of Munich}, \orgaddress{\city{Munich}, \country{Germany}}}
\affil[4]{\orgdiv{Department of Geography}, \orgname{National University of Singapore}, \orgaddress{\city{Singapore}, \country{Singapore}}}


\abstract{
The shift toward high-quality urbanization has brought increased attention to the issue of ``urban villages'', which has become a prominent social problem in China. However, there is a lack of available geospatial data on urban villages, making it crucial to prioritize urban village mapping. In order to assess the current progress in urban village mapping and identify challenges and future directions, we have conducted a comprehensive review, which to the best of our knowledge is the first of its kind in this field. Our review begins by providing a clear context for urban villages and elaborating the method for literature review, then summarizes the study areas, data sources, and approaches used for urban village mapping in China. We also address the challenges and future directions for further research. 
Through thorough investigation, we find that current studies only cover very limited study areas and periods and lack sufficient investigation into the scalability, transferability, and interpretability of identification approaches due to the challenges in concept fuzziness and variances, spatial heterogeneity and variances of urban villages, and data availability. Future research can complement and further the current research in the following potential directions in order to achieve large-area mapping across the whole nation: 1) establish a unified standard of urban villages in China to accommodate significant variances and fuzzy concepts, 2) address the challenges of data availability via flexible use of available multisource data and explore potential use of data-driven image super-resolution approaches, 3) create public benchmarks to ensure fair comparison and focus on the scalability, transferability, and interpretability of urban village recognition approaches, 4) initiate a crowdsourcing program to enable effective and efficient data collection and validation as well as application.
This review not only supports urban village-related research in China, but also contributes valuable knowledge from a Chinese perspective to global informal settlements mapping research and the achievement of the United Nations' Sustainable Development Goals (SDGs). (Updated review at \url{https://github.com/rui-research/urban-village-review})
}



\maketitle

\section{Introduction}\label{sec:introduction}
Over the past 40 years of rapid urbanization since the reform and opening up, a large number of urban villages have formed in large cities in China. 
The urban village, or ``Cheng Zhong Cun'' (literally means villages in cities) in Chinese, is an unplanned and densely populated area with relatively small scale, lagging infrastructure, and unsatisfactory sanitation in the city \citep{tong2009Hum.Geogr.,liuUrbanVillagesChina2010}. 

The formation of urban villages in China is a complexity of rapid urbanization and unique institutional regulations. In China, the urban and rural areas and residents have been institutionally divided into two distinct groups, which is referred to as the urban-rural dual structure. In terms of property rights, urban villages have special legal form, as they fall outside urban administration, and building activities are managed by the village collective, which dictates how the land is transformed and utilized \citep{sa2023urban}. This land management system leads to the formation of ‘islands’ or ‘vacuum’ within cities which lead to the high-density and unplanned developments of urban villages disrespecting urban regulation.

Due to inadequate planning and management, urban villages have caused imbalances in urban infrastructure, challenges in social governance, and a negative urban image \citep{zhengUrbanVillagesChina2009,tong2009Hum.Geogr.,liuUrbanVillagesChina2010,gao2021segregation}.
Despite the drawbacks, urban villages act as sanctuaries for a substantial number of migrant workers in cities. They represent the initial stepping stone for many young people making their way in urban areas, offering crucial housing and fundamental life support for the low-income population \citep{tong2020migrant}. 
Thus, while urban villages present significant social governance challenges and contribute to urban dysfunctions, they also serve as important hubs for providing basic living necessities to low-income groups and newcomers to the city. This dual role underscores the complexity of addressing issues related to urban villages and highlights the need for comprehensive strategies that balance social welfare concerns with urban development imperatives.

In recent years, the Chinese government is putting significant emphasis on the redevelopment of urban villages which is a key component contributing to high-quality urban development \citep{Ye2024CityPlanningReview}.
Comprehensive and detailed data on the spatial distribution of urban villages are an important basis for urban renewal and urban village transformation. However, there is a serious data shortage at present, especially for large-area mapping. The traditional data collection of urban villages relies on field survey and manual collection, which has the problems of long period, labor-consuming, and slow update \citep{zhengUrbanVillagesChina2009,chenHierarchicalApproachFinegrained2022}. Therefore, the study of automated algorithms for identifying urban villages is of great significance to the efficient acquisition of information on urban low utility land and urban renewal.
In addition, despite of significant differences in terms of residential environment, spatial form, and spatial composition, urban villages can be regarded as a special form of urban informal settlements in Chinese context \citep{huangSpatiotemporalDetectionAnalysis2015,taubenbockMorphologyArrivalCity2018}.

Compared to informal settlements in non-Chinese contexts, research on identifying urban villages in Chinese cities is limited and faces significant challenges, particularly with large-area mapping. This review aims to bridge this gap by consolidating current research on Chinese urban village mapping, summarizing both progress and challenges. Our primary goal is to assess the current state of research and identify major challenges in large-area urban village mapping. This effort will contribute to global informal settlement mapping and support the United Nations' Sustainable Development Goals (SDGs) from a Chinese perspective.
The major contributions of the article can be summarized as follows.
\begin{itemize}
    \item To the best of our knowledge, this is the first comprehensive review summarizing the current research progress on urban village mapping in China. This topic is a crucial component of global informal settlement mapping and SDG 11, which however have received limited attention in previous studies.
    \item In this review, we clearly summarize and clarify key concepts and relevant terminology, which is a crucial step often overlooked in the literature.
    \item We systematically summarize and critically review the study areas, data sources, and methodologies employed in urban village mapping research, establishing a solid foundation for understanding the current progress in this field.
    \item Building on this comprehensive overview of existing research, we identify significant challenges and potential pathways for large-area urban village mapping, highlighting future research directions.
\end{itemize}

The rest of the paper is organized as follows. Section \ref{sec:context} defines the context of urban villages surveyed in this review. Section \ref{sec:review} describes the methodology about how the systematic literature review has been carried out. Section \ref{sec:data} summarizes the study areas of current research as well as the data sources used. Section \ref{sec:approaches} reviews the approaches used for urban village mapping. Section \ref{sec:challenges} points out the current challenges and future opportunities for large-area urban village mapping. Section \ref{sec:conclusion} concludes the paper.

\section{Terminology and context of ``urban villages''}\label{sec:context}
The term ``urban village'' per se has different connotations in different contexts, such as ethnic enclaves, new models of community development, and informal settlements \citep{wangUrbanVillageGlobal2022}. Particularly, in this review, we mainly focus on the concept of informal settlements within the rapid growing metropolises in the Chinese context.

Broadly speaking, urban villages can be regarded as a special form of urban informal settlements in the context of China, as they both indicate urban residential areas which are out of the formal planning and associated with overcrowded, low-income residents; however, strictly speaking, urban villages are very different from the kind of informal settlements we usually refer to (such as slums) \citep{kufferSlumsSpace152016,kufferScopeEarthObservationImprove2018,mahabirCriticalReviewHigh2018}. Informal settlements grow as a consequence of cities, while urban villages existed before cities and have been engulfed in the spatial expansion process. Specifically, the formation of urban villages in China is a complexity resulting from the combination of the unique Chinese institutions and rapid urbanization \citep{tong2011Geogr.Res.}.
Urban villages are islands or vacuums of urban regions with special regulations and lax enforcement which eventually lead to the unique urban fabric and spatial form distinguishing them from informal settlements in non-Chinese context \citep{huangSpatiotemporalDetectionAnalysis2015,taubenbockMorphologyArrivalCity2018}.

It should also be noted that, in addition to urban informal settlements \citep{fanFineScaleUrbanInformal2022}, other relevant terms can also be found in literature from time to time to denote urban villages in China, such as migrant housing \citep{doleire-oltmannsObjectBasedClassificationApproach2011}, unplanned areas \citep{crivellariSuperresolutionGANsUpscaling2023}, and deprived areas \citep{abascalDomainsDeprivationFramework2022}. Although with subtle semantic differences and focuses, they usually refer to the same object. To facilitate discussion, we will use the term ``urban village'' throughout this review.
In addition, in the Chinese context, there are also terms with similar and confusing meanings, such as shantytown \citep{heURBANSHANTYTOWN2018,du2020large}. While urban villages are informal residential areas within cities, shantytowns emphasize the poverty and harsh living conditions of informal settlements. Urban villages may have more infrastructure and services compared to shantytowns and are typically more integrated with the urban fabric of the city. The similarity and differences further adds to the ambiguity and difficulty of urban village identification.
Detailed comparison between these terminologies can be found in Table \ref{tab:terms}.

\begin{table*}[!htbp]
    \centering
    \caption{Comparison of similar terminologies. Note that these terms are often ambiguous and may overlap in meaning, each emphasizing different aspects.}
    \label{tab:terms}
{\resizebox{\linewidth}{!}{
    \begin{tabular}{p{3cm}p{12cm}} \hline 
         Term&  Meaning \\\hline 
         Informal settlement&Informal settlements lack adequate access to potable water, basic sanitation, electricity, and other essential infrastructure. Housing in these areas is often characterized by poor structural quality, high density, and insecure ownership \citep{un-habitat2004}.\\ \hline 
         Slum&Slums are defined as areas with insufficient access to safe water, inadequate sanitation, and other critical infrastructure. They are marked by overcrowded living conditions, poor housing quality, and insecure residential status. Moreover, slums are frequently overlooked by public authorities, failing to be recognized as integral parts of the city \citep{un-habitat2004}.\\ \hline 
         Migrant housing&Migrant housing refers to accommodations for the large number of rural migrants who move to cities in search of employment. In China, the household registration system restricts rural migrants' access to essential social welfare benefits, including housing. As a result, social housing tailored to the needs of this population is referred to as migrant housing  \citep{wu2004sources}.\\ \hline 
         Unplanned areas&Unplanned areas pertain to the planning context and are defined as regions that lack detailed plans or land subdivision guidelines, and do not comply with existing planning and building regulations \citep{khalifa2011redefining}.\\ \hline 
         Deprived areas&Deprived areas pose social, environmental, and ecological risks to health and well-being. They often lack legal access to land, essential social amenities such as schools and healthcare centers, and basic infrastructure like roads and sewage systems \citep{thomson2020need}.\\ \hline 
         Shantytown&The term is associated with poor physical and socioeconomic conditions. In China, it is widely used in government policies to describe dilapidated housing or illegally constructed shanties found in older inner-city communities or rundown villages in (sub)urban and rural areas \citep{li2018shantytown}.\\ \hline
    \end{tabular}
}}
\end{table*}

\begin{figure}[!htbp]
    \centering
    \includegraphics[width=\linewidth]{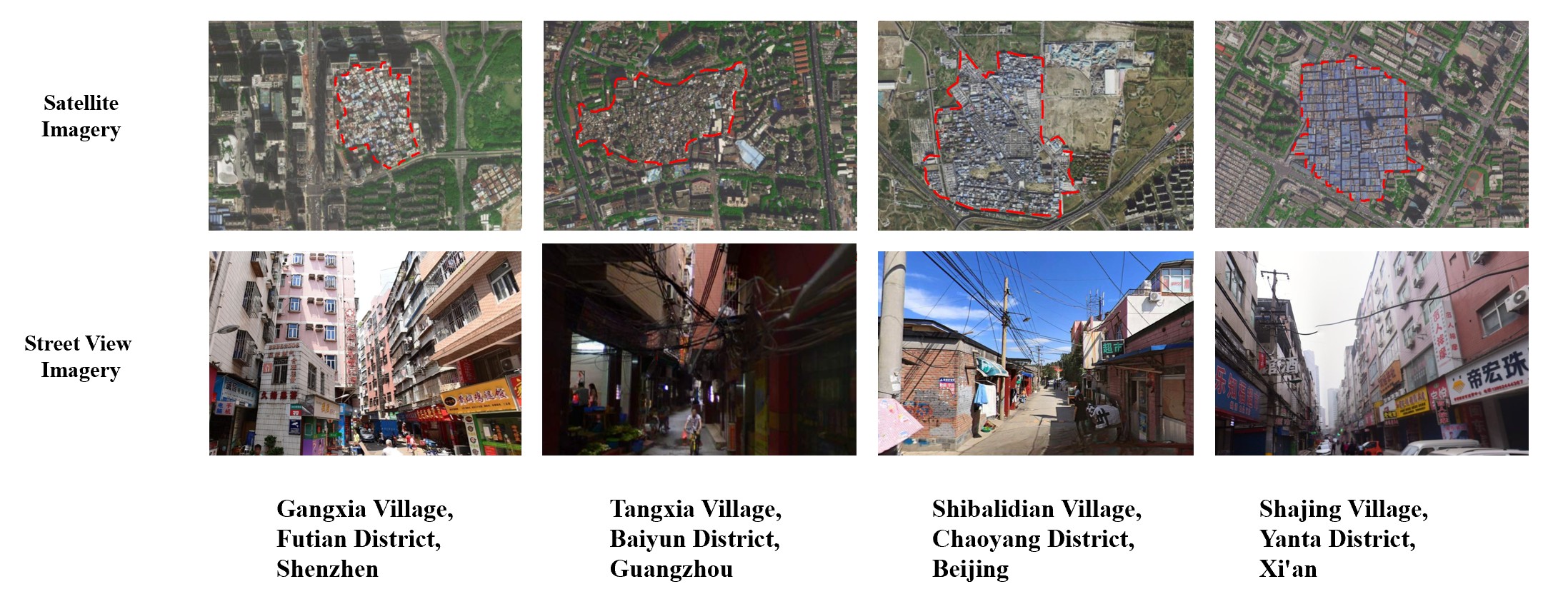}
    \caption{Typical examples of urban villages in four mega-cities in China. Both the satellite and street view images are sourced from Baidu Maps (2023). Boundaries of the urban villages in these cities, identified based on government documents and field surveys, are marked with red dashed lines in the satellite images.}
    \label{fig:uv-examples}
\end{figure}

China's vast size and diverse geography, urban landscapes, history, and culture result in urban villages exhibiting complex, varied physical and geographic patterns, with significant differences across regions and cities, as illustrated in Figure \ref{fig:uv-examples}.
The satellite and street view imagery of urban villages in cities such as Shenzhen, Guangzhou, Beijing, and Xi'an vividly illustrates these differences. For example, Shenzhen and Guangzhou showcase densely packed buildings and narrow streets indicative of high population densities and limited space, while Beijing displays a more dispersed arrangement with low-rise buildings, typical of northern rural architectural styles, and wider, more open streets. In Xi'an, the urban villages are characterized by densely packed but orderly building layouts; however, the streets remain narrow and bustling with activity, posing potential fire safety and other hazards. This diversity in urban village forms and layouts across the country highlights the complexities and challenges involved in urban planning and management.
The variation and diversity further pose great challenges for urban village detection, especially for large-area mapping. 

\section{Review Methodology}\label{sec:review}

Based on the context of urban villages, we conducted a systematic search and review of the literature related to urban village mapping in China. 
We used keywords (both in English and Chinese) including ``urban village'' (or related terms listed in Table \ref{tab:terms}); ``identification'', ``detection'', or ``mapping''; ``China''; ``remote sensing'' or ``geographic information science/system (GIS)'' to retrieve articles from academic databases including Web of Science, China National Knowledge Infrastructure (CNKI), and Google Scholar, covering literature published up to May 2024. Given our goal to understand the methods for mapping and identifying urban villages, we excluded purely theoretical or case studies that do not focus on the identification of urban villages. 
Additionally, to clarify the current research areas and data sources for mapping urban villages, we excluded articles that did not explicitly indicate the study area, or presented doubtful data quality and experiment results.

By adopting this approach, our study provides a comprehensive overview of the progress in urban village mapping in China. We conducted two rounds of relevance screening: an initial abstract screening, followed by a full-text screening to remove papers not relevant to our review scope. Through these two rounds, we identified a final corpus of 28 papers for inclusion in our review and synthesis, comprising 22 in English and 6 in Chinese. The earliest urban village identification work in our search results dates back to 2011. According to the publication venues, we categorized these studies into four research fields: Urban Studies, GIS and Remote Sensing, Artificial Intelligence (AI) and Big Data, and Agricultural Studies.

\begin{figure}[!htbp]
    \centering
    \includegraphics[width=1\linewidth]{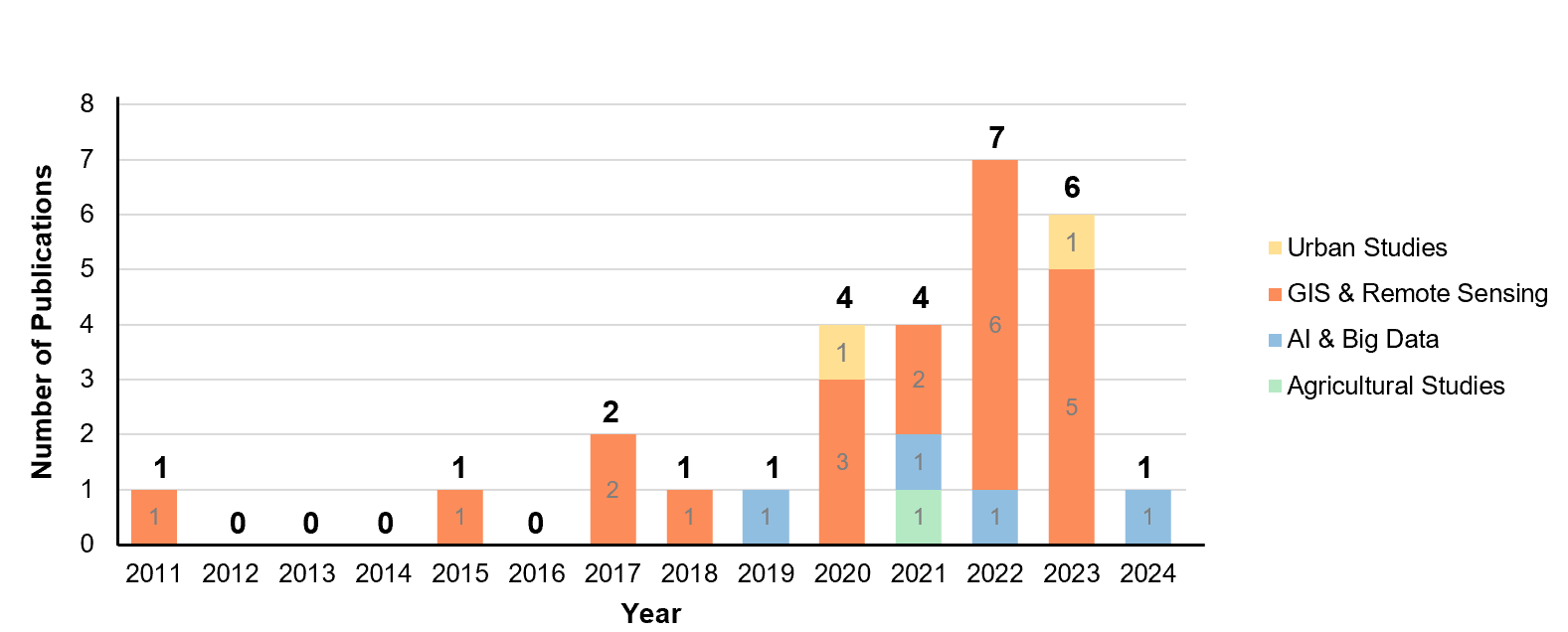}
    \caption{Number of publications per year related to urban village mapping in China. The publication venues have been categorized into four classes according to research fields, i.e., Urban Studies, GIS and Remote Sensing, AI and Big Data, and Agricultural Studies. (The annual number of publications reflects data collected up to May 2024.)}
    \label{fig:publication}
\end{figure}

The current status and trend of publication is shown in Figure \ref{fig:publication}.
As can be seen, there are very limited research of urban village mapping in China before 2020. Starting from 2020, there has been a noticeable uptrend in the volume of research publications, culminating in a peak in 2022. This surge is predominantly observed in the field of GIS and Remote Sensing, highlighting an escalating academic and technological dedication to urban village mapping. This focus is crucial given the complex and varied physical spaces and geographic landscapes of urban villages across different regions, as previously discussed. Furthermore, contributions from Computer Science conferences, particularly in the domains of AI and Big Data, have consistently been made over the years, reflecting an enduring interest of the community in themes that intersect with Urban Studies. 
However, despite this overall growth in research output, publications addressing fields like Agricultural Studies and particular aspects of Urban Studies continue to be intermittent, pointing to a persistent gap in targeted research efforts. This gap underscores the necessity for a more focused approach in understanding and addressing the unique challenges presented by urban villages, enhancing the efficacy of large-area mapping initiatives.

\section{Study areas and data for urban village mapping}\label{sec:data}
\subsection{Studied cities}
\begin{figure}[!htbp]
    \centering
    \includegraphics[width=0.98\linewidth]{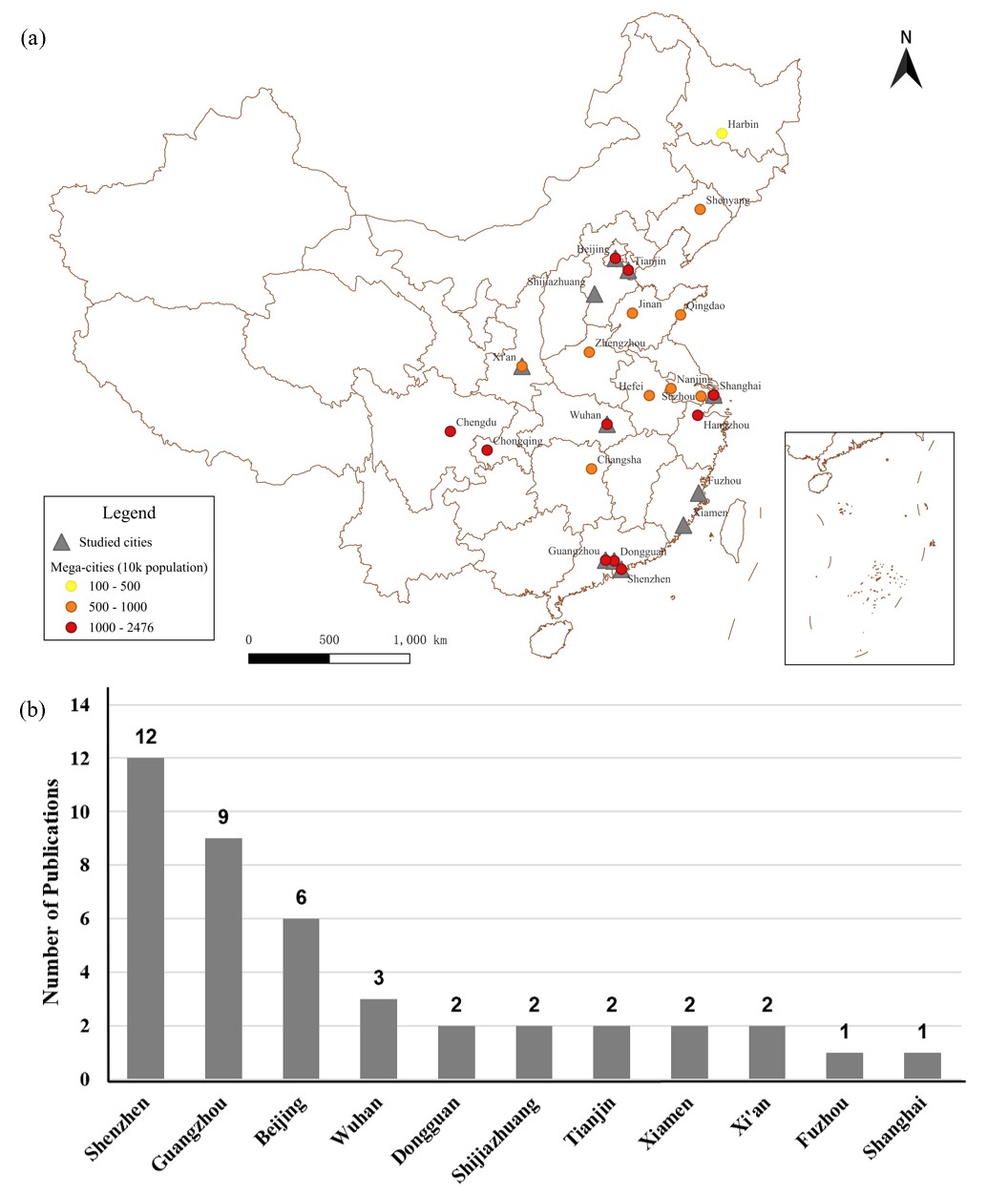}
    \caption{Statistics of publications from geographic perspective. (a) Spatial distribution of studied cities and top 20 most populous mega-cities in China. (Note: Mega-cities are divided according to the resident population of urban areas, and the data comes from the 2022 Urban Construction Statistics Yearbook of the Ministry of Housing and Urban-Rural Development in China).
    (b) Number of publications for studied cities. (Note: Research may encompass either entire cities or specific areas within cities.)
}
    \label{fig:study-areas}
\end{figure}

According to the reviewed publications, the research areas only cover 11 cities (as shown in Figure \ref{fig:study-areas}), and the vast majority of research is concentrated in a few limited cities, as shown in Figure \ref{fig:study-areas}(b), such as Shenzhen \citep{huangSpatiotemporalDetectionAnalysis2015,liUnsupervisedDeepFeature2017,liuUseLandscapeMetrics2017,mastMappingUrbanVillages2020,chenHierarchicalApproachFinegrained2022,fanMultilevelSpatialChannelFeature2022,huangComprehensiveUrbanSpace2023,chen2023Geomat.Inf.Sci.WuhanUniv.}, Guangzhou \citep{doleire-oltmannsObjectBasedClassificationApproach2011,shiDomainAdaptionFineGrained2020,panDeepLearningSegmentation2020,cui2022Natl.RemoteSens.Bull.,zhang2021ActaGeod.Cartogr.Sin.,feng2021RS4NatRes}, and Beijing \citep{chenMultimodalFusionSatellite2022,weiGaofen2SatelliteImagebased2023,xiaoContextualMasterSlaveFramework2023,feng2021Trans.Chi.Soc.Agr.Mach.}. 
In comparison, as shown in Figure \ref{fig:study-areas}(a), there are 20 mega-cities in China with resident population of urban areas over 5 millions \citep{mohurd2023}.
This research status quo that focuses on a very few mega-cities and large cities is understandable mainly due to two reasons. First, there is a general bias of urban studies to focus on large cities due to the significance of urban challenges as well as resources and policy inclination \citep{tjarks2024seeing}. Second, as urban villages formed as a consequence of rapid urbanization and unique institutional background, they usually concentrate in these dense urban areas. However, it should be noted that China has over 100 large cities with a resident population of urban areas more than 1 million \citep{mohurd2023}. This shows that the current research coverage area cannot meet the demand for spatial data on urban villages over a large area.
In addition, research on the entire city is relatively scarce in the cities studied \citep{chenHierarchicalApproachFinegrained2022,fanMultilevelSpatialChannelFeature2022,huangComprehensiveUrbanSpace2023}, and the majority of research only covers limited areas of the city, such as part of the administrative region or the area covered by remote sensing image scenes. 

\subsection{Data sources}
Due to the inherent complexity of urban villages, a variety of data sources have been exploited for urban village mapping in China, which have been summarized in Table \ref{tab:data}.
\begin{table*}[!htbp]
    \centering
\caption{Data types and sources for urban village identification.}
\label{tab:data}
{\resizebox{\linewidth}{!}{
    \begin{tabular}{cp{3cm}p{5cm}c} \hline 
         Category&  Data&  Source& Spatial Form  \\ \hline 
         Remote sensing&  Optical / Multispectral imagery&  Gaofen-1/2 \citep{crivellariSuperresolutionGANsUpscaling2023,weiGaofen2SatelliteImagebased2023,weiVanishingRenewalLandscape2023,cui2022Natl.RemoteSens.Bull.,feng2021RS4NatRes}, QuickBird \citep{huangSpatiotemporalDetectionAnalysis2015,liUnsupervisedDeepFeature2017,liuUseLandscapeMetrics2017,doleire-oltmannsObjectBasedClassificationApproach2011}, WorldView-2 \citep{huangSpatiotemporalDetectionAnalysis2015,liuUseLandscapeMetrics2017,panDeepLearningSegmentation2020}, GeoEye-1 \citep{huangSpatiotemporalDetectionAnalysis2015,liuUseLandscapeMetrics2017}, SPOT-5 \citep{doleire-oltmannsObjectBasedClassificationApproach2011,chen2023Geomat.Inf.Sci.WuhanUniv.,chenHierarchicalApproachFinegrained2022}; Google Earth \citep{chenIdentifyingUrbanVillages2019,shiDomainAdaptionFineGrained2020,mastMappingUrbanVillages2020,huangComprehensiveUrbanSpace2023,chenMultimodalFusionSatellite2022,feng2021Trans.Chi.Soc.Agr.Mach.,fanMultilevelSpatialChannelFeature2022,fanFineScaleUrbanInformal2022,fanUrbanInformalSettlements2022}, ArcGIS \citep{zhangUVSAMAdaptingSegment2024}, Tianditu \citep{zhang2021ActaGeod.Cartogr.Sin.}, Baidu Map \citep{xiaoContextualMasterSlaveFramework2023}; Sentinel-2 \citep{crivellariSuperresolutionGANsUpscaling2023}, Landsat \citep{doleire-oltmannsObjectBasedClassificationApproach2011}&  Raster grids \\ \hline 
         &  SAR imagery&  TerraSAR-X \citep{weiMonitoringUrbanVillages2015}&  Raster grids \\ \hline 
 GIS \& Geospatial big data& Street view images& Baidu Map \citep{cui2022Natl.RemoteSens.Bull.,fanMultilevelSpatialChannelFeature2022}, Tencent Map \citep{huangComprehensiveUrbanSpace2023,chenMultimodalFusionSatellite2022,fanMultilevelSpatialChannelFeature2022}& Points \\ \hline 
 & POI& Baidu Map \citep{niuIntegratingMultipleData2020,xiaoContextualMasterSlaveFramework2023,zhao2018Geogr.Geo-Inf.Sci.}, Gaode Map \citep{liMappingUrbanVillages2023,chen2023Geomat.Inf.Sci.WuhanUniv.}& Points \\ \hline 
 & Road networks& OpenStreetMap \citep{chenUVLensUrbanVillage2021,weiVanishingRenewalLandscape2023,zhang2021ActaGeod.Cartogr.Sin.,feng2021RS4NatRes}& Polylines \\ \hline  
         &  Building footprint&  Baidu Map \citep{fanFineScaleUrbanInformal2022,zhao2018Geogr.Geo-Inf.Sci.}&  Polygons \\ \hline 
 Mobility data& Taxi trajectories& Government \citep{chenUVLensUrbanVillage2021,huangComprehensiveUrbanSpace2023,chen2023Geomat.Inf.Sci.WuhanUniv.}& Spatio-temporal points \\ \hline 
 & Shared bike trajectories& Mobike \citep{chenUVLensUrbanVillage2021,niuIntegratingMultipleData2020}& Spatio-temporal points \\ \hline 
 & Mobile phone positioning data& Tencent user density map \citep{zhao2018Geogr.Geo-Inf.Sci.,fanUrbanInformalSettlements2022}& Raster grids \\ \hline
    \end{tabular}
    }}
\end{table*}

Similar to slum mapping in non-Chinese context, current research on the identification of urban villages mainly focuses on intelligent interpretation of very-high-resolution (VHR) remote sensing images, which are usually with spatial resolution lower than 3 m. Specifically, the data are from optical or multi-spectral remote sensing imagery, and there are mainly two kinds of sources. The first data source is VHR satellite imagery including Gaofen-1 \citep{weiVanishingRenewalLandscape2023,weiMonitoringUrbanVillages2015,feng2021RS4NatRes} (2 m), Gaofen-2 \citep{weiGaofen2SatelliteImagebased2023,cui2022Natl.RemoteSens.Bull.} (1 m), QuickBird \citep{huangSpatiotemporalDetectionAnalysis2015,liUnsupervisedDeepFeature2017,liuUseLandscapeMetrics2017} (2.4 m), WorldView-2 \citep{huangSpatiotemporalDetectionAnalysis2015,liuUseLandscapeMetrics2017} (2 m), GeoEye-1 \citep{huangSpatiotemporalDetectionAnalysis2015,liuUseLandscapeMetrics2017} (2 m), SPOT-5 \citep{chenHierarchicalApproachFinegrained2022,doleire-oltmannsObjectBasedClassificationApproach2011,chen2023Geomat.Inf.Sci.WuhanUniv.} (2.5 m). The advantage of VHR satellite imagery is the high quality, and rich information of multi-spectral bands besides visible bands (RGB). In addition, time-series satellite images usually have better consistency and good for spatio-temporal detection and analysis \citep{huangSpatiotemporalDetectionAnalysis2015,liuUseLandscapeMetrics2017}.
The other data source is online map services, such as Google Earth \citep{chenIdentifyingUrbanVillages2019,chenUVLensUrbanVillage2021,shiDomainAdaptionFineGrained2020,mastMappingUrbanVillages2020,huangComprehensiveUrbanSpace2023,chenMultimodalFusionSatellite2022,fanFineScaleUrbanInformal2022,fanMultilevelSpatialChannelFeature2022,fanUrbanInformalSettlements2022,feng2021Trans.Chi.Soc.Agr.Mach.}, ArcGIS online \citep{zhangUVSAMAdaptingSegment2024}, Baidu Map \citep{xiaoContextualMasterSlaveFramework2023}, and Tianditu \citep{zhang2021ActaGeod.Cartogr.Sin.}. 
For example, the Google Earth has high-resolution remote sensing imagery with global coverage and relatively reliable 2D positional accuracy, which can be used for a large number of applications including slum mapping \citep{potereHorizontalPositionalAccuracy2008,yuGoogleEarthVirtual2012,kufferSlumsSpace152016}.
These online sources have both advantages and disadvantages. The advantage of online map services is the easy accessibility of the VHR remote sensing imagery, which often can be acquired free of charge. However, there are also disadvantages. The first disadvantage is the quality of the data, which suffer from distortion and inconsistent preprocessing. The second disadvantage is that these images usually only include visible spectral bands of RGB, without other spectral information. In addition, there are usually only the latest imagery available and the time-series images are often more difficult to acquire from online map sources.

In addition to VHR remote sensing images, there are also exploration of freely available multi-spectral Sentinel-2 imagery, with medium spatial resolution of 10m, for slum mapping and urban village detection.
Sentinel-2 imagery have been used for slum mapping in Mumbai, India \citep{wurmExploitationTexturalMorphological2017,vermaTransferLearningApproach2019,pengMappingSlumsMumbai2023}.
It is also explored for urban village mapping after upscaling  via image super-resolution method \citep{crivellariSuperresolutionGANsUpscaling2023}.
Landsat imagery with spatial resolution of 30 m is used as auxiliary source for urban village detection in coarse scale \citep{doleire-oltmannsObjectBasedClassificationApproach2011}.
Apart from multispectral imagery, SAR imagery has also been used for slum mapping and urban village mapping. 
The polarimetric SAR imagery is demonstrated to be useful in slum mapping \citep{wurmSlumMappingPolarimetric2017}.
Following this line, SAR imagery such as TerraSAR-X is also tested for urban village monitoring combining with multi-spectral imagery \citep{weiMonitoringUrbanVillages2015}.

The intricate internal configurations and diverse spatial forms of urban villages present significant challenges for their identification through single remote sensing imagery alone. Consequently, a variety of additional data sources are utilized to enhance urban village mapping, often in conjunction with remote sensing imagery. These include street view images \citep{chenMultimodalFusionSatellite2022,fanMultilevelSpatialChannelFeature2022,huangComprehensiveUrbanSpace2023,cui2022Natl.RemoteSens.Bull.}, points-of-interest (POI) data \citep{chenHierarchicalApproachFinegrained2022,xiaoContextualMasterSlaveFramework2023,liMappingUrbanVillages2023,feng2021RS4NatRes}, road networks \citep{chenUVLensUrbanVillage2021}, building footprints \citep{fanFineScaleUrbanInformal2022}, vehicle trajectory data \citep{chenUVLensUrbanVillage2021,chenHierarchicalApproachFinegrained2022,huangComprehensiveUrbanSpace2023}, and mobile positioning data \citep{fanUrbanInformalSettlements2022}. Each data source offers unique insights but also comes with its own set of limitations.
Street view images provide detailed, ground-level perspectives absent in top-down remote sensing imagery, capturing vertical building information and the immediate environment. However, their coverage is limited to areas accessible via roads, and they are not universally available through mapping services, further constrained by spatiotemporal limitations \citep{caoIntegratingSatelliteStreetlevel2023}. POI data, offering semantic details about locations, are more accessible and have broader spatial coverage but suffer from data quality issues and are typically restricted to the most current information \citep{chenHierarchicalApproachFinegrained2022}. Road networks and building footprints contribute valuable constraints for mapping but vary in accessibility and data quality from open sources like OpenStreetMap \footnote{\url{https://www.openstreetmap.org}}, particularly in less developed cities \citep{herfortSpatiotemporalAnalysisInvestigating2023}. Spatiotemporal data sources like vehicle trajectories and mobile positioning data are valuable for understanding human mobility and activities and crucial for distinguishing urban villages \citep{chenHierarchicalApproachFinegrained2022}. Yet, these data types are challenging to acquire due to privacy concerns, often available only at coarse resolutions and at a high cost.
While integrating multiple data sources significantly improves identification accuracy, collecting them, particularly mobility and time-series data, is not straightforward \citep{chenHierarchicalApproachFinegrained2022,fanUrbanInformalSettlements2022,huangComprehensiveUrbanSpace2023}. This data integration can negatively impact the scalability of mapping methods across larger spatial areas or over extended time periods, due to difficulties in data collection and the limitations inherent to each data type.

\subsubsection{Ground-truth data}
Ground-truth data of urban villages are crucial for effective data-driven model training and validation for urban village identification. 
In general, the public geospatial data of urban villages in China is lacking. Although few municipal governments have made the urban village map public in their urban planning documents, such as Shenzhen \citep{sz2019planning} and Guangzhou \citep{gz2024planning}, but without proper geospatial reference and scale, which are difficult to reproduce for practical use.
Therefore, to create ground-truth data for model training and algorithms validation, normally, ground-truth data are annotated by researchers themselves or by volunteers they recruited through field surveys or remote sensing image interpretation.
However, manually labeling data for model training and validation can be highly time-consuming and labor-intensive.
Currently, there are very limited public available annotated urban village datasets. 
S$^2$UV dataset \citep{chenMultimodalFusionSatellite2022} consists of paired satellite and street view images of urban villages as well as their labels collected in Beijing, Tianjin, and Shijiazhuang. 
UV-SAM dataset \citep{zhangUVSAMAdaptingSegment2024} is composed of satellite imagery and associated mask images (indicating the boundaries of urban villages) of Beijing and Xi'an. 
However, these datasets doesn't include public available geospatial coordinates.
In addition, there are over 2000 building footprints within urban villages annotated but without any geospatial reference as well \citep{liu2020training}.
The lacking of geospatial data of urban villages further shows the need and urgency for urban village mapping research.

\section{Approaches for urban village mapping}\label{sec:approaches}

\rev{
In this section, we review and summarize the approaches for urban village mapping based on two key dimensions: (1) the input data (i.e., remote sensing-based vs. multisource data-based) and (2) the underlying methodological philosophies (i.e., classification-oriented vs. segmentation-oriented). While ‘classification-oriented’ and ‘segmentation-oriented’ paradigms may exhibit technical overlaps, they embody fundamentally different perspectives in spatial pattern recognition:
\begin{itemize}
\item \textbf{Classification-oriented approaches} are object-oriented, typically following a two-step workflow: first delineating basic study units (e.g., objects, parcels, grids) and then classifying them. In this paradigm, features of the study units are usually extracted and classified independently, with limited consideration of spatial relationships between units. The delineation of study units is critical as it determines the effective spatial resolution of the analysis. These methods are particularly suitable for integrating multisource data with varying spatial resolutions and accuracies, as well as for scenarios involving predefined study units.
\item \textbf{Segmentation-oriented approaches} adopt a pixel-based perspective, referring to semantic/instance segmentation techniques that directly partition images at the pixel level while modeling spatial context through neighboring pixel relationships. These approaches leverage automatic feature learning via deep neural networks in an end-to-end workflow, seamlessly integrating feature extraction and classification. This paradigm generates coherent regions rather than independent study units, making it suitable for applications that require fine-grained boundary delineation.
\end{itemize}
In addition, we summarize the performance evaluation metrics for both types of approaches following the review of methods.
}

\subsection{\rev{Remote sensing-based approaches}}
\rev{Currently, most research on urban village mapping in China relies solely on remote sensing images, with a focus on intelligent interpretation using VHR remote sensing images.}

\subsubsection{Classification-oriented approaches}
Earlier research on urban village mapping has primarily relied on remote sensing images and utilized classification-oriented approaches, with object-based and scene-based methods being the most prevalent. 
Object-Based Image Analysis (OBIA) \citep{blaschkeGeographicObjectBasedImage2014} is commonly employed for mapping urban villages. For instance, \cite{doleire-oltmannsObjectBasedClassificationApproach2011} presents an OBIA-based method that utilizes various sources of satellite images for this purpose.
In contrast, scene-based classification focuses on larger areas, typically encompassing several hundred meters. \cite{huangSpatiotemporalDetectionAnalysis2015} introduces a scene-based semantic feature that outperforms traditional Bag of Visual Words (BoVW) and Linear Discriminant Analysis (LDA) features, subsequently applying Support Vector Machines (SVM) and Random Forests (RF) for scene-level classification. 
Additionally, \cite{liUnsupervisedDeepFeature2017} proposes an unsupervised deep feature learning method to extract deep features from scene-level images, using SVM for classification. Similarly, \cite{liuUseLandscapeMetrics2017} employs landscape metrics and transfer learning for classifying scene images, again utilizing RF and SVM for the classification process.

\subsubsection{Segmentation-oriented approaches}
With the advancement of deep learning, end-to-end image segmentation has become more feasible and effective, leading to an increasing number of studies on urban village mapping that employ deep neural network-based segmentation models. These studies frame urban village mapping as an image segmentation challenge using remote sensing images, leveraging deep neural networks for both semantic and instance segmentation as well as boundary extraction.
For example, \cite{chenIdentifyingUrbanVillages2019} approaches urban village mapping as an instance segmentation problem, utilizing Mask R-CNN \citep{he2017mask} to segment urban village areas on Xiamen Island using VHR Google Earth imagery. Similarly, \cite{mastMappingUrbanVillages2020} employs a fully convolutional network (FCN) \citep{long2015fully} to map urban village areas in Shenzhen based on Google Earth imagery. 
Further advancing this field, \cite{weiGaofen2SatelliteImagebased2023} compares various fully convolutional neural networks, including FCN \citep{long2015fully}, UNet \citep{ronneberger2015unet}, and ResUNet \citep{zhang2018road}, to segment urban village areas from Gaofen-2 satellite imagery. Additionally, \cite{panDeepLearningSegmentation2020} utilizes UNet \citep{ronneberger2015unet} to analyze WorldView-2 satellite imagery for detecting and classifying buildings within urban villages in Guangzhou.

Beyond fully convolutional neural networks, \cite{zhangUVSAMAdaptingSegment2024} adapts the latest Segment Anything Model (SAM) \citep{kirillov2023segment} for specialized urban village segmentation. Some research explicitly addresses the transferability of methods; for instance, \cite{shiDomainAdaptionFineGrained2020} considers urban village mapping as a semantic segmentation problem and employs deep adversarial domain adaptation techniques to enhance segmentation results across Shenzhen and Guangzhou.
Emerging studies are also beginning to explore the use of freely available, lower-resolution satellite imagery. Notably, \cite{crivellariSuperresolutionGANsUpscaling2023} employs super-resolution generative adversarial networks (SR-GANs) to upscale Sentinel-2 imagery, subsequently addressing it as a semantic segmentation problem using ResUNet.

\subsubsection{Summary}
The process of urban village detection using remote sensing images has evolved significantly. Initially, it relied on handcrafted features and traditional machine learning algorithms for image classification. These early methods required the delineation of study units first, which limited the resolution of the results and often resulted in lower accuracy. In contrast, modern deep learning approaches automate feature extraction, enabling the handling of complex images and scenes, thereby improving segmentation and classification accuracy. However, these advanced models face challenges, particularly regarding explainability and potential overfitting in limited areas. Future research should focus on enhancing the interpretability as well as the scalability and transferability of these models. Additionally, urban village detection results often suffer from fragmentation, necessitating further post-processing steps, such as image morphology operations \citep{chenHierarchicalApproachFinegrained2022}. Furthermore, super-resolution methods show promise for utilizing lower-resolution yet accessible remote sensing imagery for large-area and long-period urban village mapping \citep{crivellariSuperresolutionGANsUpscaling2023}.

\subsection{\rev{Multisource data-based approaches}}
Due to the complex internal structure of urban villages and significant differences in spatial form and external landscape, a single remote sensing image often cannot effectively identify them. Therefore, research on the identification of urban villages based on multisource data fusion (including street view images \citep{chenMultimodalFusionSatellite2022,fanMultilevelSpatialChannelFeature2022,huangComprehensiveUrbanSpace2023}, POI data \citep{chenHierarchicalApproachFinegrained2022,xiaoContextualMasterSlaveFramework2023}, road network data \citep{chenUVLensUrbanVillage2021}, building footprint data \citep{fanFineScaleUrbanInformal2022}, vehicle trajectory data \citep{chenUVLensUrbanVillage2021,chenHierarchicalApproachFinegrained2022,huangComprehensiveUrbanSpace2023}, mobile positioning data \citep{fanUrbanInformalSettlements2022}, etc.) has gradually attracted attention recently.

\subsubsection{Classification-oriented approaches}
For data sources beyond remote sensing images, classification-based approaches are often preferred due to the typically coarser spatial resolutions and varying levels of accuracy. This makes it more natural to integrate these data sources into study units with coarser resolutions. Various scales of study units have been explored, including pixel \citep{chenHierarchicalApproachFinegrained2022}, object \citep{doleire-oltmannsObjectBasedClassificationApproach2011}, building \citep{niuIntegratingMultipleData2020}, grid \citep{huangComprehensiveUrbanSpace2023}, land parcel \citep{fanFineScaleUrbanInformal2022}, and scene \citep{huangSpatiotemporalDetectionAnalysis2015}.

For instance, \cite{chenHierarchicalApproachFinegrained2022} introduces a coarse-to-fine hierarchical recognition method for pixel-level urban village classification, utilizing SPOT-5 satellite imagery, POI, and taxi trajectories. The method first segments the satellite imagery using an OBIA approach to identify potential urban village objects, followed by pixel-level classification based on these identified objects. Both pixel and object classifications extract handcrafted features and deep features from the satellite imagery, along with handcrafted features derived from POI and taxi trajectories. 

At the grid level, \cite{huangComprehensiveUrbanSpace2023} proposes a deep learning-based approach that integrates satellite imagery, street view images, and taxi GPS trajectories to identify urban villages. Similarly, \cite{chenMultimodalFusionSatellite2022} presents a dual-branch deep neural network that integrates remote sensing and street view images at the grid level. Several studies focus on land parcel-level analysis as well, exploring deep neural network-based fusion methods to integrate satellite images with street view images \citep{fanMultilevelSpatialChannelFeature2022}, time-series human activity data \citep{fanUrbanInformalSettlements2022}, and building footprint data \citep{fanFineScaleUrbanInformal2022} for urban village mapping.
\cite{weiVanishingRenewalLandscape2023} maps urban villages in Beijing using Gaofen-1 satellite imagery, OSM road networks, and a random forest classifier, focusing on block-level analysis, which is similar to land parcel studies.

While most studies treat study units as independent entities, some research seeks to enhance classification performance by leveraging the inter-correlation among study units. For instance, \cite{xiaoContextualMasterSlaveFramework2023} proposes a graph neural network-based approach to integrate remote sensing imagery and POI data for urban village detection at the grid level. Furthermore, in addition to integrating remote sensing images with other data sources, some studies explore methods that do not rely on remote sensing imagery at all. \cite{liMappingUrbanVillages2023} proposes using deep neural networks and POI data alone to map urban villages at the grid level.

\subsubsection{Segmentation-oriented approaches}
While classification-based approaches are commonly used, they are not the only option. Some research focuses on integrating multiple data sources for urban village mapping using segmentation-oriented methods. For instance, \cite{chenUVLensUrbanVillage2021} combines remote sensing imagery, road networks, and shared bike trajectories to define urban village boundaries and subsequently predict population distribution in Xiamen and Shanghai. This study employs a two-stage process for urban village mapping: the first stage delineates coarse land parcel boundaries using road networks and bike trajectories, while the second stage utilizes Mask R-CNN to segment remote sensing imagery, guided by the initial land parcel boundaries.

\subsubsection{Summary}
The identification of urban villages through the fusion of multisource data typically employs a classification-based approach. This process usually begins with the division of spatial units to accommodate data with varying spatial resolutions. The subsequent steps closely resemble those of classification methods based on remote sensing images, with the added requirement of extracting and fusing features from the multisource data. 
Alternatively, one can view the additional data as a prior constraint, allowing for semantic segmentation of remote sensing images within that framework. While the former approach is limited by the resolution of the study units, the latter assumes that the additional data is sufficiently accurate. Generally, multisource data fusion approaches face transferability challenges due to the difficulties in obtaining diverse datasets, particularly mobility data that raises privacy concerns. This issue becomes more pronounced when analyzing large areas over extended time periods. Future research may focus on leveraging multisource data for coarse detection or targeting particularly challenging areas \citep{chenHierarchicalApproachFinegrained2022}.

\subsection{Performance evaluation}
For performance evaluation of urban village mapping, the evaluation metrics can be categorized into two types, i.e., classification-based and segmentation-based.

For classification-oriented approaches, the classification metrics are used for performance evaluation, such as overall classification accuracy (CA), detection accuracy (precision, recall, F1-score), Kappa coefficient ($\kappa$). 

\begin{equation}
    CA = \frac{\left|TP\right| + \left|TN\right|}{\left|TP\right| + \left|TN\right| + \left|FP\right| + \left|FN\right|}
\end{equation}

\begin{equation}
    precision = \frac{\left|TP\right|}{\left|TP\right| + \left|FP\right|},~
    recall = \frac{\left|TP\right|}{\left|TP\right| + \left|FN\right|},~
    \textrm{F1-score} = \frac{2 \cdot precision \cdot recall}{precision + recall}
\end{equation}

\begin{equation}
    \kappa = \frac{p_o-p_e}{1-p_e}
\end{equation}
where $\left|TP\right|$ and $\left|TN\right|$ are the numbers of urban villages and non-urban villages that are correctly classified respectively, while $\left|FP\right|$ and $\left|FN\right|$ are the numbers of wrongly classified urban villages and non-urban villages respectively.
In Kappa coefficient, $p_o$ is the relative observed agreement among raters (the same as the classification accuracy $CA$), which is the proportion of agreement between the classification results and the ground truth data; $p_e$ is the hypothetical probability of chance agreement, which is calculated based on the marginal totals of the confusion matrix.
The Kappa coefficient ranges from -1 to 1. A higher Kappa coefficient indicates better agreement between classification results and ground truth data, while a lower coefficient suggests poorer agreement.

For segmentation-oriented approaches, segmentation accuracy is often used, such as intersection over union (IoU). 

\begin{equation}
    IoU = \frac{\left|\{\textrm{ground-truth~pixels}\} \cap \{\textrm{detected~pixels}\}\right|}{\left|\{\textrm{ground-truth~pixels}\} \cup \{\textrm{detected~pixels}\}\right|}
\end{equation}

As current research is conducted in different study areas and with various study units, and even with different evaluation metrics, the urban village mapping performances often cannot be directly compared across studies. 
For instance, taking Shenzhen and Beijing as examples, there is significant variation in the performance of urban village detection. 
In studies using remote sensing imagery alone, reported classification accuracies of classification-oriented approaches in Shenzhen range from 79\% \citep{huangSpatiotemporalDetectionAnalysis2015} to over 98\% \citep{liUnsupervisedDeepFeature2017,chen2023Geomat.Inf.Sci.WuhanUniv.}, while in Beijing the reported results range from 82\% \citep{liMappingUrbanVillages2023} to 95\% \citep{weiVanishingRenewalLandscape2023}. For segmentation-oriented approaches using IoU as an evaluation metric, the reported performances in Shenzhen range from 68\% \citep{shiDomainAdaptionFineGrained2020} to over 85\% \citep{fanFineScaleUrbanInformal2022}. In contrast, in Beijing, the IoU is reported at around 72\% \citep{zhangUVSAMAdaptingSegment2024}.
Additionally, within the same city, different areas or different study units can lead to significant variations in detection performance. This on the one hand highlights the complexity of urban village monitoring and the challenge of adapting detection methods to varied urban landscapes and data conditions, on the other hand also accentuates the current need for research of large-area mapping.

For methods that integrate multiple sources of data, reported performances generally are improved noticeably. For example, in Shenzhen, combining street view image data with high-resolution remote sensing imagery has increased classification accuracy by over 4\%  compared using remote sensing imagery alone \citep{fanMultilevelSpatialChannelFeature2022}. Furthermore, integrating Tencent population density data with high-resolution remote sensing imagery has increased the overall accuracy by 10\% \citep{fanUrbanInformalSettlements2022}. 
This indicates that multisource data fusion can significantly enhance the recognition performance of urban villages.

It is also of note that current evaluation metrics all take the identification task as hard boundary detection, which ignores the ambiguity and fuzziness of the boundaries of urban villages (as discussed in Section \ref{sec:challenge-ambiguity}). It is worth to evaluate the performance taking the uncertainty into consideration in the future regarding the inherent ambiguity of the task.

\section{Challenges and future directions for large-area urban village mapping}\label{sec:challenges}
Overall, compared to informal settlements mapping in non-Chinese context, current research on the identification of urban villages in Chinese cities is relatively scarce. To facilitate future research, we identify the challenges and point out the future directions for large-area urban village mapping.

\subsection{Contextual ambiguity and fuzzy boundaries}\label{sec:challenge-ambiguity}
As mentioned in Section \ref{sec:context}, ``urban villages" have different connotations in various contexts, and they exhibit diverse landscapes and morphology. Current research targets are varied and lack a unified definition for ``urban villages", which poses a significant challenge for large-area urban village mapping in China. A detailed comparison of ``urban village" and terms with similar contextual meanings is presented in Table \ref{tab:terms}, which helps to alleviate contextual ambiguity issues. The ambiguity in definition and context also results in unclear boundaries for urban villages. Under the current task settings, hard boundaries need to be established, which do not accurately reflect the fuzziness of the semantic connotation of urban villages. This characteristic is also shared by slums \citep{kohliUncertaintyAnalysisImage2016,pratomoCouplingUncertaintiesAccuracy2017}. 

In addition, this fuzziness also reflect on the choice of basic study units. Actually, there is a lack of uniformity in the basic spatial units and identification tasks of research. Due to differences in understanding of urban villages, research areas, available data, etc., the basic spatial units for urban village identification research are not uniform, and there are various studies based on pixel \citep{chenHierarchicalApproachFinegrained2022}, object \citep{doleire-oltmannsObjectBasedClassificationApproach2011}, building \citep{niuIntegratingMultipleData2020}, grid \citep{huangComprehensiveUrbanSpace2023}, land parcel \citep{fanFineScaleUrbanInformal2022}, or scene \citep{huangSpatiotemporalDetectionAnalysis2015}, and other different scales, and the urban village identification tasks are correspondingly defined as segmentation or classification problems. 
Different spatial units offer distinct advantages and disadvantages. Pixel-based classification provides high resolution and detail but may be limited by computational demands and sensitivity to noise. Object-based classification enhances accuracy by incorporating spatial relationships and reducing noise but is more complex and dependent on segmentation quality. Building-based classification excels in urban areas by accurately identifying structures, though it is less useful in rural settings and requires high-resolution data. Grid-based classification offers a balanced approach by organizing data into regular cells, but it may lose detail and is sensitive to grid size. Parcel-based classification is effective for land use analysis based on predefined boundaries but can be constrained by the accuracy of these boundaries. Scene-based classification provides a broad view of large areas, suitable for large-area patterns but may sacrifice detail and demand significant computational resources. In general, selecting the right spatial unit depends on the classification objectives, data characteristics, and the needed detail and accuracy.

Consequently, directly comparing methods and results across various studies of urban village mapping presents significant challenges. These difficulties stem from the need to address the variance and ambiguity inherent in the contexts, boundaries, and study units of urban villages. To facilitate comprehensive mapping of urban villages over large areas, it is imperative for future research to precisely define what constitutes an urban village. Additionally, there should be a clear documentation of the methodologies used for delineating boundaries and selecting study units. This approach will ensure that benchmarking, comparison, and reproducibility are conducted fairly and effectively.

\subsection{Data availability}
Although remote sensing imagery can cover large areas, it is not easy to acquire high-quality VHR remote sensing images, especially for large cities. Covering an entire metropolis requires multiple scenes of costly VHR imagery, making it expensive to obtain. This issue is even more pronounced for time-series analysis of urban villages, which requires multi-temporal imagery. This is also part of the reason why many research only focus on part of the regions within a city \citep{huangSpatiotemporalDetectionAnalysis2015}.
There are several cost-effective alternative data sources for VHR satellite imagery. The first choice is using online map services such as Google Earth, which has very-high-resolution remote sensing imagery with almost global coverage and relatively reliable 2D positional accuracy and has been extensively used for applications such as slum mapping \citep{potereHorizontalPositionalAccuracy2008,yuGoogleEarthVirtual2012,kufferSlumsSpace152016}. However, these data sources also suffer from data quality issues, such as distortion and inconsistent preprocessing.
Another choice is trying to exploit freely available satellite imagery with relatively lower spatial resolution. For example, although Sentinel-2 imagery data are with the spatial resolution of 10 m, they have been explored and used for slum mapping in Mumbai, India \citep{wurmExploitationTexturalMorphological2017,vermaTransferLearningApproach2019,pengMappingSlumsMumbai2023}. In addition, the Sentinel-2 imagery has also been tested with super-resolution methods to enhance its spatial resolution for urban village mapping, which has been demonstrated to be effective \citep{crivellariSuperresolutionGANsUpscaling2023}. As the data-driven image super-resolution methods advance, it becomes increasingly promising \citep{wang2020deep,wang2022comprehensive}.

Other sources of data, such as POIs, street view images, vehicle trajectories, mobile phone positioning data, often provide social and economic attribute information and human activity information on the basis of geographical landscape information provided by remote sensing images, helping to enhance the identification ability of urban villages. However, these kinds of data usually are more difficult to acquire than remote sensing images, and they usually suffer from data sparsity and quality issues. In addition, time-series data are even more difficult to acquire.
To accommodate the sparsity of the data sources, one way is to use hierarchical methods \citep{chenHierarchicalApproachFinegrained2022}.
To address the data missing issues, data imputation for multisource data fusion might be a viable way \citep{caoDeepLearningbasedRemote2020}.

While the limitations of the various data sources have been discussed in Section \ref{sec:data},  there are more issues in the context of large-area mapping.
Street view images provide invaluable visual context but are costly and have uneven coverage across different areas, with updates not always keeping pace with urban development. Their use is often restricted by licensing agreements, making broad-scale access and up-to-date imagery a challenge.
POIs and road networks are critical for understanding the functional layout of urban villages. They are generally accessible through open-source platforms, offering a cost-effective solution with reasonable quality. However, the accuracy of POIs can vary, and while road networks usually offer comprehensive coverage, the frequency of updates might not fully capture rapidly evolving urban landscapes.
Building footprints are essential for spatial analysis in urban village mapping, providing moderate to high-quality data depending on source imagery. The main challenges lie in the variable coverage and the infrequency of updates, which may not reflect recent developments, though they are moderately accessible from both open-source and proprietary sources.
Vehicle trajectories and mobile positioning data deliver high-quality, granular insights into mobility patterns, crucial for urban planning and traffic management. Despite their value, the high cost, variable coverage, and significant privacy concerns limit their accessibility, making them less feasible for widespread use in large-scale urban village mapping projects.

In consolidating data for large-area urban village mapping, the balance between availability, cost, quality, coverage, and granularity becomes pivotal. Each data type's characteristics must be carefully considered to effectively capture the complex nature of urban village environments.
While integrating diverse data sources for urban village mapping across all the study areas can be challenging and oftentimes infeasible, especially in the context of large-area mapping; one promising future direction is to leverage these sources to identify urban villages in difficult-to-map regions, which could enhance model performance. Additionally, these data sources can serve as reference for cross-checking information in more typical areas, helping to label high-quality urban village data for training and validation purposes.

\subsection{Scalability and transferability of approaches}

Currently, there is insufficient attention paid to the scalability and transferability of urban village mapping methods both in space and time dimensions in order to expand the proposed approaches to cover larger areas and longer time periods.
In the context of urban village detection algorithms, scalability refers to the algorithm's ability to handle growing data volumes and complexity efficiently, while transferability denotes its capacity to be adapted effectively to different geographic regions or urban environments. Together, these qualities ensure that the algorithm remains both robust and versatile across various contexts and data scales.

As mentioned in Section \ref{sec:data}, due to the limitations of current research within a single city or a few cities, the performance evaluation of identification methods often cannot well reflect the transferability across regions. On the other hand, as the identification of urban villages is still in a relatively early stage, there is relatively limited research beyond accuracy evaluation to account for transferability explicitly \citep{shiDomainAdaptionFineGrained2020}.
There are also rare research about the various landscape and morphology of urban villages \citep{liuUseLandscapeMetrics2017}, which is highly likely to be connected to model transferability according to the patterns exhibited by slums \citep{taubenbockMorphologyArrivalCity2018}.
Furthermore, in the context of multisource data-based approaches, it is essential to consider additional data sources beyond remote sensing images; however, acquiring these additional data sources can be even more challenging, posing a barrier to the scalability and transferability of these approaches. As a result, it becomes difficult to compare current methods on an equal and direct basis, highlighting the necessity for benchmarks to facilitate fair comparisons.

Besides in space, there is also limited research extended in temporal dimension. For remote sensing-based approaches, the spatial-temporal analysis of urban villages requires for time-series satellite imagery data and also demands for transferable methods across data of different time periods \citep{huangSpatiotemporalDetectionAnalysis2015,liuUseLandscapeMetrics2017,weiVanishingRenewalLandscape2023}.
In addition, it is challenging to detect changes through time-series data per se \citep{shi2020change,wen2021change}, the situation is even more pronounced for urban village mapping since there are consensus issues regarding the boundaries.
As for multisource data-based approaches, data sources such as street view images, mobility data, and building information, are not applicable to all cities, and acquiring multi-temporal data is even more challenging, which further limits their scalability and transferability for temporal analysis.

\subsection{Potential applications and associated challenges}
Timely and accurate urban village maps are crucial for understanding the social dynamics of cities and promoting sustainable urban development. These maps can be analyzed independently or alongside other spatial data, such as population density and infrastructure provision. This combined analysis reveals the spatio-temporal profiles of urban villages from multiple perspectives.
On the spatial dimension, examining the spatial distribution of urban villages reveals various patterns across different regions. For instance, analyzing rank-size distributions or landscape indices can highlight the regional heterogeneity in the spatial patterns of urban villages \citep{friesenSimilarSizeSlums2018}. On the temporal dimension, analyzing dynamic maps of urban villages to reveal the trends of urban village growth is helpful for understanding the urban dynamic mechanism in the context of urbanization. These analyses allow urban designers and planners to have a better understanding of urban villages and formulate a holistic strategy to transform urban villages, thereby improving the living conditions of urban populations experiencing poverty.

Demographic profiles of urban villages represent another valuable application of these maps. By integrating spatial demographic data with urban village maps, planners and policymakers can gain deeper insights into the socioeconomic characteristics of urban village residents. This combined analysis can reveal disparities in demographic profiles such as age, gender, income, education, and employment, which in turn can inform targeted interventions to address inequality and promote inclusive urban development. The advancement of information and communication technologies has led to the growing availability of extensive spatial demographic data, including open data sources (e.g., WorldPop \citep{tatem2017worldpop}, LandScan \citep{dobson2000landscan}), mobile phone positioning data \citep{deville2014dynamic}, government official census data, etc. However, these existing datasets generally lack specific background information on urban villages. Urban village maps can help enrich these datasets, enhancing the analysis of demographic profiles within urban villages and supporting more informed and effective urban planning.

The main challenges in applying urban village maps primarily involve issues of compatibility. Compatibility can be influenced by two key factors: ambiguity and spatial resolution. Firstly, ambiguity in definition and context, as discussed in Section \ref{sec:challenge-ambiguity}, can lead to the misuse of urban village maps across various domains. For example, some urban village redevelopment projects in urban planning focus solely on buildings, while others consider urban villages as communities, including surrounding public spaces and infrastructure. Users of urban village maps must ensure that the data's definition aligns with their project's needs and carefully select the appropriate spatial units. Secondly, adequate spatial resolution is crucial for effective application. Urban villages typically encompass neighborhood-scale spaces, requiring detailed spatial information for urban planning case studies. Additionally, spatial resolution affects the compatibility with other spatial data. A high-resolution urban village map can be easily downsampled to match other datasets with any spatial scales. However, excessively detailed spatial information can raise ethical concerns, which will be discussed in the next section (Section \ref{sec:challenge-ethical}).

\subsection{Ethical issues}\label{sec:challenge-ethical}
Urban villages normally reflect the places where migrant urban residents with relatively low income gathered \citep{zhengUrbanVillagesChina2009,gao2018exploring}. Thus, there is a risk of ethical issues of revealing the precise locations of urban villages, and it may lead to privacy issues, stigmatization, security issues, and even evictions of migrants \citep{kufferSlumsSpace152016}.
Therefore, cautiousness is needed to publish the urban village map data to the public. There are several strategies to navigate the potential risks. 
The first is to add certain constraints to the data accessibility, such as providing data on request and only with the consent of certain promise of no violation of data abuse. 
Another strategy is to deliberately decrease the accuracy of published data by techniques such as aggregation to coarse scales, for example, using grids of spatial resolution of several hundred meters or aggregating to county-level administrative regions.
The strategies can also be combined, with coarse data publicly available and fine-grained data on request.
In a word, we need to handle these urban village map data carefully to prevent good intentions from leading to negative outcomes.

\subsection{Summary}

In the context of large-area urban village mapping, the aforementioned challenges in contextual definition, available data, identification approaches, downstream applications, as well as ethical issues take on specific dimensions and importance that are specific to the task and require considerable efforts to address them in the future work.

In addition to these specific challenges and associated research directions, there are also general challenges highly relevant with the task.
The cost and processing time become significant considerations given the vast areas to be covered and the detailed analysis required. The validation of model outcomes is crucial in this setting, as inaccuracies can lead to misrepresentation of communities, potentially affecting urban planning and policy decisions. Fieldwork, although challenging, becomes indispensable for validating the results obtained through remote sensing and AI models. Initiating a crowdsourcing program to enable effective and efficient data collection and validation can be a promising direction.
Moreover, the uncertainties and interpretation of models also pose a particular challenge in the context of urban village mapping. Urban villages are complex socio-economic entities, and their representation through models requires careful consideration of various factors and indicators. Communicating these model results to end-users or stakeholders, including urban planners, local governments, and the residents of the urban villages themselves, is vital. 

On the other hand, the opportunities provided by recent advancements in cloud computing, deep learning, geospatial processing software, and explainable AI algorithms \citep{caoResponsibleUrbanIntelligence2023} are particularly relevant in tackling the complexities of mapping large-area urban villages. These technological advancements can help in accurately identifying and delineating urban village boundaries, understanding their spatial patterns, and monitoring their change over time.

The success of large-area urban village mapping projects depends not only on the accuracy of the maps produced but also on how these maps are used to inform urban development and improve the lives of urban village residents. Thus, addressing these challenges and leveraging the available opportunities is crucial for the advancement of large-area urban village mapping endeavors.

\section{Conclusion}\label{sec:conclusion}
With the shift to high-quality urbanization, ``urban villages'' have become an increasingly prominent social problem. Despite the significance of studying urban villages, there are scarce spatial urban village data available. Thus, it is of high importance for urban village mapping.
To investigate the latest progress made in urban village mapping and point out challenges and future directions, we make a comprehensive review, which to the best of our knowledge is the first in this field. 
Our thorough investigation reveals that current studies on urban village mapping are limited in terms of study areas and periods, and they insufficiently address the scalability, transferability, and interpretability of identification approaches. This shortfall arises from challenges such as concept fuzziness, spatial heterogeneity, and data availability. To advance research and facilitate large-area urban village mapping nationwide, future efforts can focus on the following potential directions: 1) establishing a unified standard for urban villages in China to accommodate significant variances and fuzzy concepts; 2) tackling data availability challenges by flexibly utilizing available multisource data and exploring the potential of data-driven image super-resolution techniques; 3) creating public benchmarks to ensure fair comparisons while emphasizing the scalability, transferability, and interpretability of urban village recognition methods; and 4) initiating a crowdsourcing program to enhance data collection, validation, and application efficiency.
The review not only facilitates urban village-related research in China, but also benefits the global informal settlements mapping research and the realization of the United Nations' Sustainable Development Goals by contributing knowledge from a Chinese perspective.

\backmatter





\bmhead{Acknowledgments}
This work is supported by the Guangzhou-HKUST(GZ) Joint Funding Program (No. 2025A03J3639), the Guangdong provincial project (No. 2023QN10H717), the National Natural Science Foundation of China (No. 42101472), Shenzhen Science and Technology Program (No. JCYJ20220818100200001), the Innovation Team of the Department of Education of Guangdong Province (No. 2024KCXTD013), and the Center for Scientific Research and Development in Higher Education Institutes, MOE, China (No. 2024HT013).

\bibliography{ref}


\begin{appendices}

\section{Abbreviations}
\begin{table}[h]
    \centering
    \begin{tabular}{p{2cm}p{8cm}} \hline  
         Abbreviation&  Full Term \\ \hline  
         AI & Artificial Intelligence \\ \hline  
         BoVW & Bag of Visual Words \\ \hline  
        CA & Classification Accuracy \\ \hline  
        FN & False Negative \\ \hline  
        FP & False Positive \\ \hline  
        GIS & Geographic Information Science/System \\ \hline  
        GPS & Global Positioning System \\ \hline  
        IoU & Intersection over Union \\ \hline  
        LDA & Latent Dirichlet Allocation \\ \hline  
        Mask R-CNN & Mask Region-Based Convolutional Neural Network \\ \hline  
        OBIA & Object-Based Image Analysis \\ \hline  
        OSM & OpenStreetMap \\ \hline  
        POI & Point of Interest \\ \hline  
        RF & Random Forest \\ \hline  
        SAM & Segment Anything Model \\ \hline  
        SDGs & Sustainable Development Goals \\ \hline  
        SR-GANs & Super-Resolution Generative Adversarial Networks \\ \hline  
        SVM & Support Vector Machine \\ \hline  
        TN & True Negative \\ \hline  
        TP & True Positive \\ \hline  
        VHR & Very-High-Resolution \\ \hline 
    \end{tabular}
    \label{tab:abbreviation}
\end{table}

\end{appendices}
\end{document}